\documentclass[11pt,titlepage]{article} 

\usepackage[utf8]{inputenc} 

\usepackage[affil-it]{authblk}

\usepackage{geometry} 
\usepackage{graphicx}
\usepackage{amsmath}

\usepackage{booktabs} 
\usepackage{array}
\usepackage{paralist} 
\usepackage{verbatim} 
\usepackage{subfig}

\usepackage{fancyhdr} 
\usepackage{sectsty}
\allsectionsfont{\sffamily\mdseries\upshape} 

\usepackage[nottoc,notlof,notlot]{tocbibind} 
\usepackage[titles,subfigure]{tocloft}

\begin{document}

\title{Direct Terrestrial Test of Lorentz Symmetry in Electrodynamics to 10$^{-18}$}
\author[1,$\#$]{Moritz Nagel}
\author[2,$\#$,*]{Stephen R. Parker}
\author[1]{Evgeny V. Kovalchuk}
\author[2]{Paul L. Stanwix}
\author[2,3]{John G. Hartnett}
\author[2]{Eugene N. Ivanov}
\author[1]{Achim Peters}
\author[2]{Michael E. Tobar}
\affil[1]{Institut f{\"u}r Physik, Humboldt-Universit{\"a}t zu Berlin, Newtonstra{\ss}e 15, 12489 Berlin, Germany}
\affil[2]{School of Physics, The University of Western Australia, 35 Stirling Highway, Crawley 6009, Australia}
\affil[3]{Institute for Photonics and Advanced Sensing, School of Physical Sciences, The University of Adelaide, Adelaide 5005, Australia}
\affil[$\#$]{These authors contributed equally to this work}
\affil[*]{stephen.parker@uwa.edu.au}

\maketitle
\abstract{Lorentz symmetry is a foundational property of modern physics, underlying the standard model of particles and general relativity. It is anticipated that these two theories are low energy approximations of a single theory that is unified and consistent at the Planck scale. Many unifying proposals allow Lorentz symmetry to be broken, with observable effects appearing at Planck-suppressed levels; thus precision tests of Lorentz invariance are needed to assess and guide theoretical efforts. Here, we use ultra-stable oscillator frequency sources to perform a modern Michelson-Morley experiment and make the most precise direct terrestrial test to date of Lorentz symmetry for the photon, constraining Lorentz violating orientation-dependent relative frequency changes $\Delta\nu$/$\nu$ to 9.2$\pm$10.7$\times10^{-19}$~(95$\%$~confidence interval). This order of magnitude improvement over previous Michelson-Morley experiments allows us to set comprehensive simultaneous bounds on nine boost and rotation anisotropies of the speed of light, finding no significant violations of Lorentz symmetry.}
\\
\\
\section{Introduction}
A significant consequence of Lorentz symmetry is the isotropic nature of the speed of light, which remains invariant under rotation and boost transformations. Measuring the isotropy of the speed of light has played an important role in physics, starting with the seminal Michelson and Morley interferometer experiment in the late 19th century~\cite{mm1}. What began as a hunt for a luminiferous ether soon evolved into tests of the revolutionary theory of special relativity. Current motivation arises from the search for hints of new physics~\cite{bluhm2006,liberati2012,tasson2014} to provide direction in the quest for a unified theory of quantum mechanics and general relativity~\cite{weinberg2009}.
\\
\\
Despite the success of unifying the weak force and electrodynamics~\cite{weinberg1967}, unification of the standard model with gravity remains elusive. Many approaches invoking string theory~\cite{ks89}, quantum gravity~\cite{myers2003,petr2009,sotiriou2009} and noncommutative field theories~\cite{noncomfield} either explicitly require or naturally permit Lorentz symmetry to be broken~\cite{weinberg2009}. Efforts to test these theories experimentally focus on searches for violations of Lorentz and Charge-Parity-Time (CPT) symmetry, which are expected to occur at the Planck scale (5.4$\times$10$^{-44}$~s, 1.6$\times$10$^{-35}$~m, 1.2$\times$10$^{19}$~GeV), with suppressed effects manifesting in regimes experimentally accessible via precision measurement~\cite{kost1995}. Many tests of Lorentz and CPT symmetry have been performed for a variety of fields, particles and forces, with no violations reported to date~\cite{datatables}. 
\\
\\
Just as the theories motivating Michelson-Morley style experiments have changed, so too has the technology used. The transition from conventional optical interferometer searches to resonator-stabilized frequency source tests~\cite{essen1955,brillet1979,micro1,micro4,Schiller2009,optical6} has enabled a rapid increase in experimental sensitivity; an overview is provided in Fig.~1. The exceptional frequency stability performance of these modern sources makes tests of Lorentz symmetry of the photon one of the most powerful experimental tools in the search for clues towards unification frameworks such as quantum gravity. 
\\
\\
Here we present the results of the most sensitive Michelson-Morley style frequency comparison experiment performed to date. We use one year of data to set new bounds on the 9 possible rotational and boost isotropies of the speed of light, with our results expressed as constraints on coefficients of the Standard Model Extension (SME)~\cite{ck2}. We find no evidence of any statistically significant violation of Lorentz symmetry of the photon.
\begin{figure}[t]
\centering
\includegraphics[width=0.8\columnwidth]{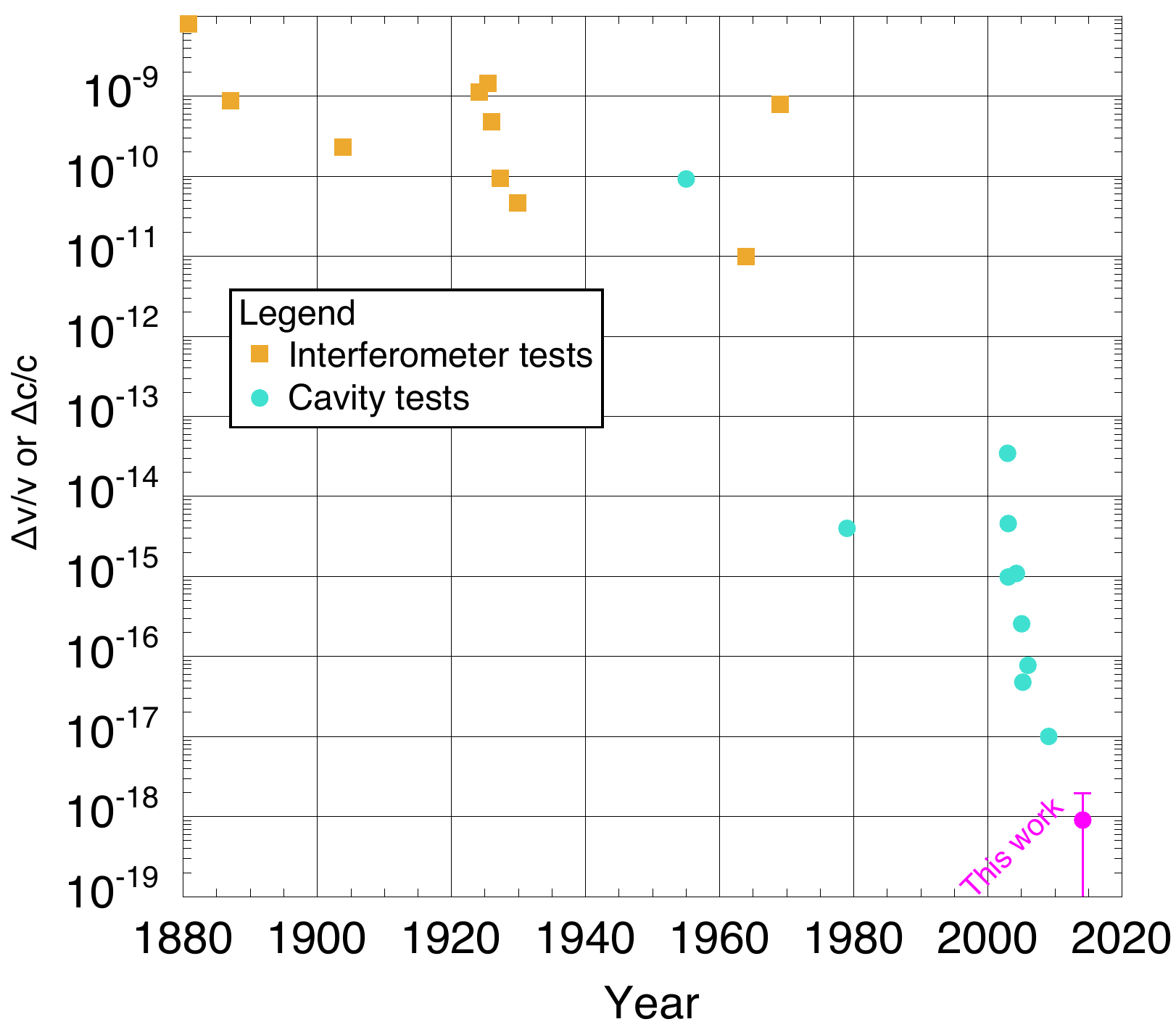}
\caption{\textbf{Historical overview of Michelson-Morley style tests of Lorentz invariance in electrodynamics.} Presented are interferometer tests (squares) that measure fractional shifts in the speed of light, $\Delta c$/$c$, and cavity-based tests (circles) such as this work that measure a fractional change in frequency, $\Delta\nu$/$\nu$. Bounds are taken directly from the results reported in original publications. A full list is provided in the Supplementary References, with numerical values given in Supplementary Table 1.}
\end{figure}
\section{Results}
\subsection{Experiment design}
A schematic of our Michelson-Morley oscillator experiment, located in Berlin (latitude 52.4$^{\circ}$), is presented in Fig.~2. Two cylindrical copper cavities were each loaded with a nominally identical cylindrical sapphire dielectric crystal. Whispering gallery modes were excited within the crystals with a resonance frequency of 12.97~GHz. Pound control electronics are used to build two loop oscillator circuits with each oscillator locked to the resonance frequency of a cavity~\cite{Locke:2008}. The cavity crystal axes were aligned perpendicular to each other such that the Poynting vectors and thus path of light propagation for the resonant modes were orthogonal to each other (see insert of Fig.~2). A fractional change in the speed of light would induce a proportional fractional change in the beat note frequency of the two oscillators. Thus, to be highly sensitive to the signals of Lorentz Invariance Violation (LIV) one needs to employ extremely low noise frequency sources. Of course, systematic noise sources also lead to frequency changes, thus we require both low noise sources and a well controlled setup. At cryogenic temperatures the frequency stability of the loop oscillator circuit was optimized, which ultimately dictated the sensitivity of the experiment (see Fig.~3).
\\
\\
We analyzed the experiment relative to a sun-centred inertial reference frame~\cite{Kostelecky:2002}. As such, passive rotation transformations were provided courtesy of Earth's daily and annual cycles. In addition, the apparatus was actively rotated in the laboratory on a tilt-stabilized high-precision air-bearing turntable that was run with a period of 100 seconds. This corresponds to the optimal performance of the experiment, which is a trade-off between oscillator frequency stability, rate of data acquisition and the influence of systematic noise. A 200 litre liquid helium reservoir allowed for 3 weeks of continuous operation before refilling was required. 
\begin{figure}[t!]
\centering
\includegraphics[width=0.7\columnwidth]{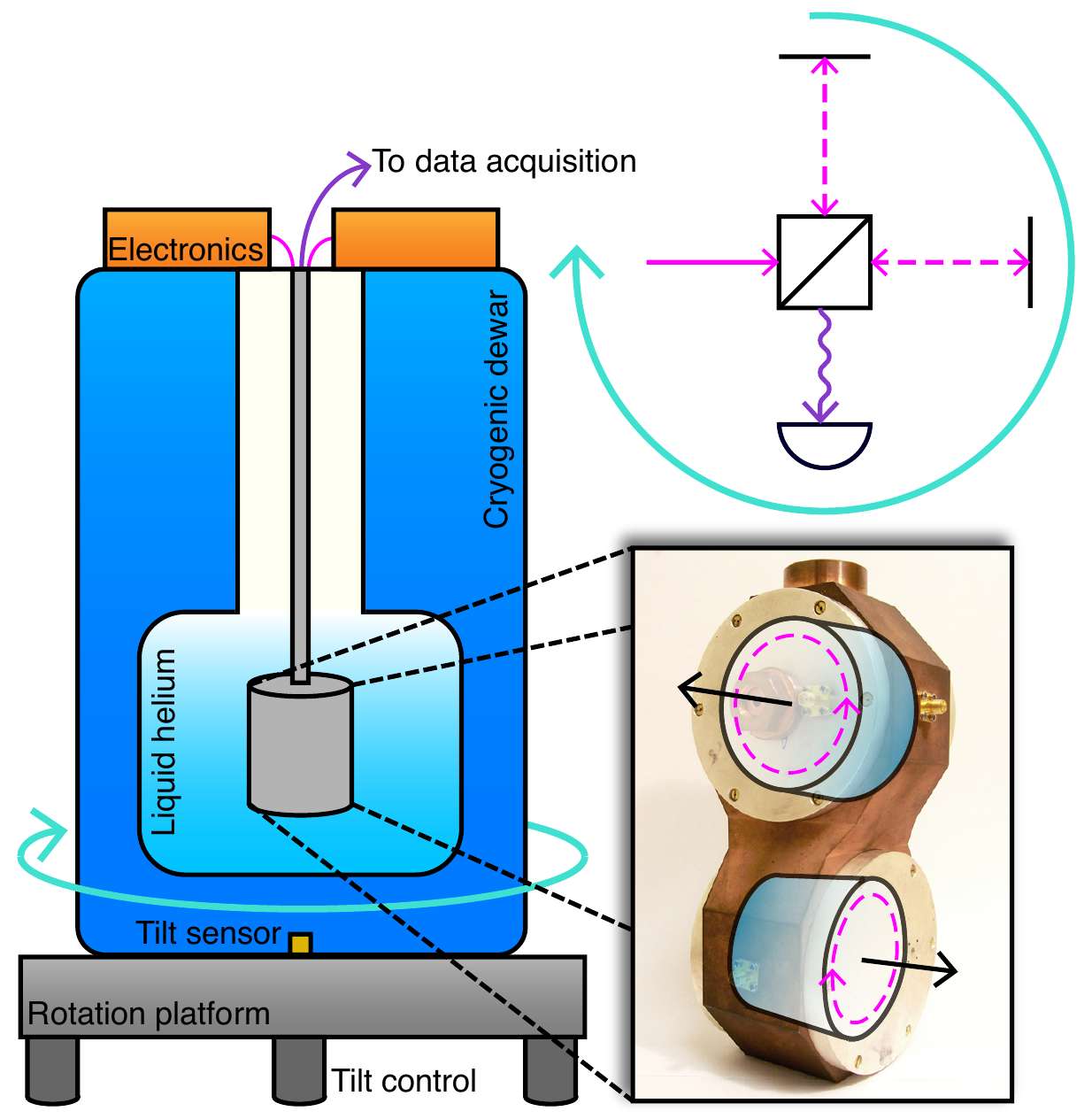}
\caption{\textbf{Schematical overview of the experimental setup.} Two sapphire cylinders were loaded in a dual-cavity mount with their crystal axes (black arrows) aligned orthogonally. This was housed within two vacuum cans and cooled to 4~K in a liquid helium dewar. A whispering gallery mode resonance was excited in each sapphire, the Poynting vector is shown by the dashed magenta arrow. Microwave and DC electronics were used to create two loop oscillators, each locked to the resonance of a sapphire. The two oscillators were beat against each other and the difference frequency was recorded. The apparatus was continuously rotated with a 100~second period on a tilt-controlled air-bearing turntable. Comparison to the original Michelson-Morley arrangement is presented in the top right to demonstrate experimental concept.}
\end{figure}
\\
\\
Due to the symmetry of the setup with respect to the sun-centered inertial reference frame, the signal of interest occurs at twice the turntable rotation frequency, 2$\omega_{\text{R}}$, with additional sideband modulations arising from Earth's sidereal rotation, $\omega_{\oplus}$, and orbit, $\Omega_{\oplus}$. This has the added benefit of suppressing the influence of any rotation induced sources of systematic noise that manifest at the fundamental turntable rotation frequency (see Fig.~3). The experimental setup is first order sensitive to LIVs of rotational transformations, with a suppressed sensitivity to symmetry breaking of boost transformations. The suppression is of order 10$^{-4}$, which is the ratio of Earth's orbital velocity to the speed of light. 
\\
\\
\subsection{Analysis}
\begin{figure}[t!]
\centering
\includegraphics[width=0.75\columnwidth]{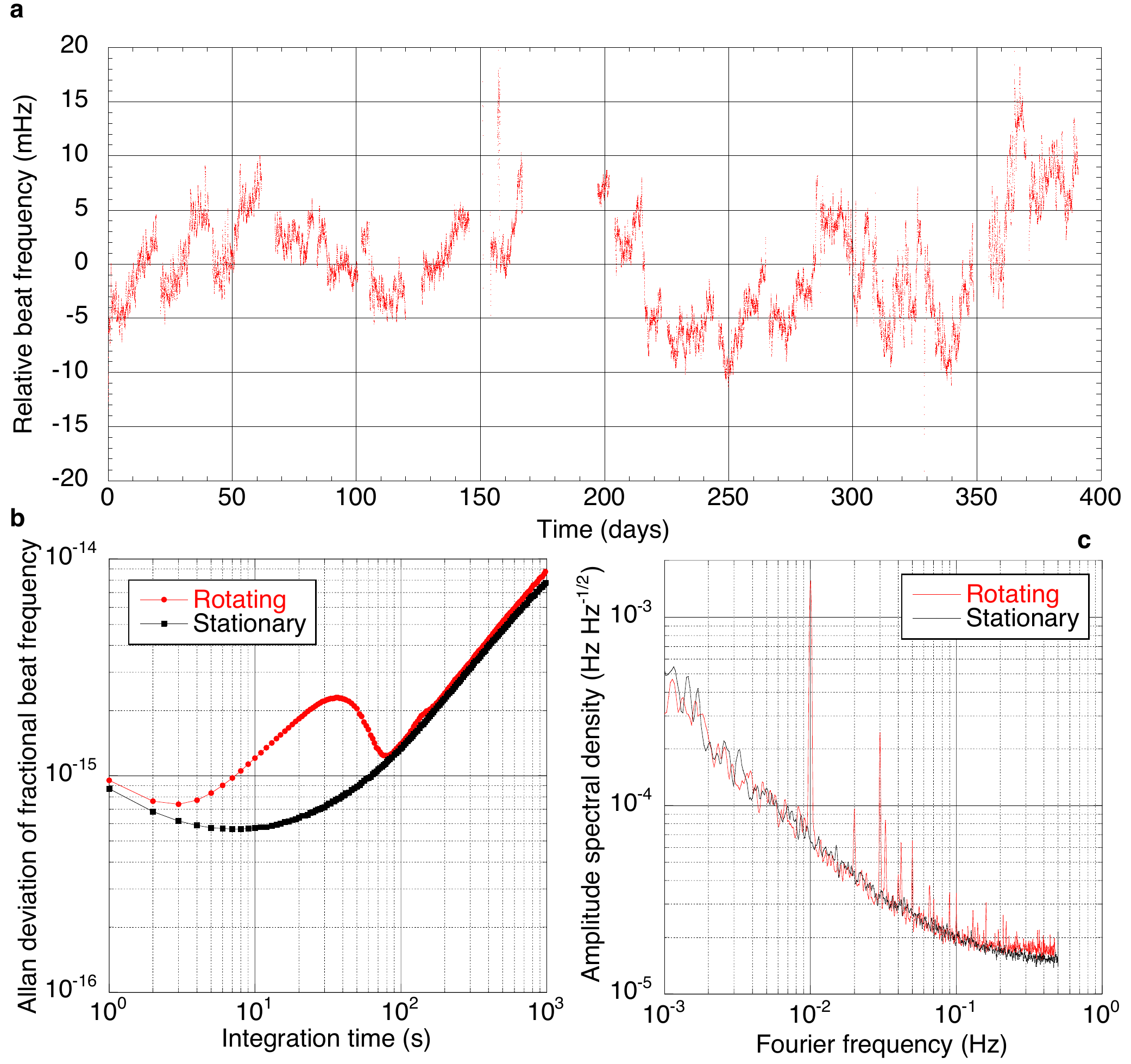}
\caption{\textbf{Beat frequency of the two cryogenic oscillators.} \textbf{a} Sampled beat frequency recorded over the course of the experiment from June 2012 to June 2013, with linear drift ($\sim$86~$\mu$Hz per day) and a 347~kHz offset subtracted for display purposes. The gap in data was due to a short period of downtime during the 2012/2013 transition. \textbf{b} Allan deviation of the fractional beat frequency, demonstrating the fundamental frequency stability of the oscillators when they are stationary in the laboratory (black trace) compared to the stability when actively rotated (red trace). Rotated stability represents typical performance and was calculated from a subset of the data taken during April 2012. \textbf{c}  Comparison of amplitude spectral density of the oscillators beat frequency when stationary (black trace) and rotated (red trace) in the laboratory, computed from the same data as in \textbf{b}. Peaks corresponding to the fundamental turntable rotation frequency and higher order harmonics can be resolved clearly, while the background noise level has not increased. Signals of interest are located at sidereal and annual sideband frequencies around the second turntable harmonic (2$\omega_{\text{R}}$=0.02~Hz), where the noise is not increased by rotation.}
\end{figure}
Data was collected over the course of a year; the beat frequency between the oscillators for the duration of the experiment is displayed in Fig.~3. The beat frequency data was analysed for periodic signals of variation corresponding to modulation frequencies of interest. Least squares regression was used to perform a fit to the fundamental turntable rotation frequency and the first harmonic,
\begin{align}
\frac{\Delta\nu_{\text{beat}}}{\nu_{\text{beat}}}=A+Bt+\sum\limits_{n=1,2} C_{n} \cos{\left(n\omega_{\text{R}}t+\phi_{n}\right)}+S_{n} \sin{\left(n\omega_{\text{R}}t+\phi_{n}\right)}. \label{eq:freq}
\end{align}
Error-weighted least squares regression was then used to fit the amplitudes $C_{n}$ and $S_{n}$ from equation~\eqref{eq:freq} to the daily variations ($\omega_{\oplus}$, 2$\omega_{\oplus}$) and finally to the annual frequencies ($\Omega_{\oplus}$, 2$\Omega_{\oplus}$). A detailed description of the data analysis process can be found in previous work~\cite{optical6,cav1,cav2} and in the methods section.
\\
\\
Taking error-weighted averages of relevant amplitudes from equation~\eqref{eq:freq} we found a 2$\omega_{\text{R}}$ amplitude of -98$\pm$6~nHz. This value of interest, 2$\omega_{\text{R}}$, is only statistically significant due to the influence of systematic noise sources (see Fig.~3), the most dominant of which is the dependency of oscillator resonance frequency on external magnetic fields, arising from the presence of impurities in the sapphire crystal~\cite{spinres} and ferrite-based microwave components. The frequency variations induced by moving the oscillators through the quasi-static magnetic field of the Earth in the laboratory are indistinguishable from a Lorentz violating signal. However, the leakage into sidereal and annual sidebands is negligible.
\\
\\
Calculating the weighted average of quadrature amplitudes for daily and twice daily variations ($\omega_{\oplus}$, 2$\omega_{\oplus}$; see Supplementary Figures~1--8) we found a frequency variation of 12$\pm$14~nHz (95$\%$ confidence interval), leading to our reported bound on the overall sensitivity of the experiment, $\Delta\nu$/$\nu$ $\le$ 9.2$\pm$10.7$\times$10$^{-19}$ (95$\%$ confidence interval). The presence of a statistically significant LIV signal in any solitary amplitude would propagate through to the final value.
\\
\\
\section{Discussion}
\begin{table}[t]
\caption[Results]{\textbf{Bounds on non-birefringent photon-sector coefficients of the minimal SME.} Errors are standard 1$\sigma$ of statistical origin. Values for $\tilde{\kappa}_{\text{e}-}$ are given in $10^{-18}$, $\tilde{\kappa}_{\text{o}+}$ in $10^{-14}$ and $\tilde{\kappa}_{\text{tr}}$ in $10^{-10}$.}
\centering
\begin{tabular}{c|c}
\hline\hline
Coefficient & Bound (Error) \\
\hline
$\tilde{\kappa}_{\text{e}-}^{XY}$ & -0.7 (1.6) \\
$\tilde{\kappa}_{\text{e}-}^{XZ}$ & -5.5 (4.0) \\
$\tilde{\kappa}_{\text{e}-}^{YZ}$ & -1.9 (3.2) \\
$\tilde{\kappa}_{\text{e}-}^{XX}-\tilde{\kappa}_{\text{e}-}^{YY}$ & -1.5 (3.4) \\
$\tilde{\kappa}_{\text{e}-}^{ZZ}$ & -286 (279) \\
$\tilde{\kappa}_{\text{o}+}^{XY}$ & -3.0 (3.4) \\
$\tilde{\kappa}_{\text{o}+}^{XZ}$ & 0.2 (1.7) \\
$\tilde{\kappa}_{\text{o}+}^{YZ}$ & -2.0 (1.6) \\
$\tilde{\kappa}_{\text{tr}}$ & -6.0 (4.0) \\
\hline\hline
\end{tabular}
\label{tab:results}
\end{table}
We use our data to place limits on coefficients of the SME~\cite{ck2}, which is an effective field theory framework containing the standard model and general relativity along with any possible Lorentz and CPT symmetry violating coefficients that could arise from the various fields. It is important to note that the SME is a purely phenomenological framework, designed to enable comprehensive searches for Lorentz and CPT symmetry violations and facilitate cataloguing and comparisons between experiments. It is convention to express results in a single inertial reference frame, in this case the widely adopted sun-centred celestial equatorial frame is used~\cite{Kostelecky:2002}.
\\
\\
It is important to point out that technically, at a fundamental level, a change in the beat frequency of the experiment is proportional to not just a change in the speed of light (due to Lorentz violation of the photon) but also to changes in the propagating medium due to Lorentz violation of the electron~\cite{electrons2003, electrons2003-2,electrons2005}. What we are actually measuring and constraining is a difference or combination of these potential effects. Due to the choice of coordinates the proton sector remains Lorentz invariant and does not contribute to any changes in the propagating medium~\cite{Kostelecky:2002}.
\\
\\
We restrict our attention to the photon sector of the minimal SME~\cite{Kostelecky:2002}, which only contains operators of renormalizable dimension in flat spacetime. The resulting possible violations of Lorentz symmetry can be divided into polarization dependent or independent effects; astrophysical constraints~\cite{bire3} have limited polarization dependent violations far beyond the reach of this work and are thus ignored. What remains are 9 coefficients constraining different effects - rotation violations, described by the 5 coefficients of the 3$\times$3 symmetric traceless matrix $\tilde{\kappa}_{\text{e}-}^{jk}$, boost violations, described by the 3 coefficients of the 3$\times$3 antisymmetric matrix $\tilde{\kappa}_{\text{o}+}^{jk}$ and an overall isotropic deviation of $c$, described by the scalar $\tilde{\kappa}_{\text{tr}}$~\cite{cav2}. Experimental sensitivities to these coefficients are provided in Supplementary Table 2.
\\
\\
For sapphire with the $c$ axis parallel to the crystal axis, as is the case in this work, the sensitivity to Lorentz violation of electrons is reduced~\cite{electrons2003-2,electrons2005}. The effect is such that the orientation coefficients, $\tilde{\kappa}_{\text{e}-}^{jk}$, and the isotropic deviation, $\tilde{\kappa}_{\text{tr}}$, should be replaced with $\tilde{\kappa}_{\text{e}-}^{jk}-2.25c^{jk}$ and $\tilde{\kappa}_{\text{tr}}+1.5c^{e}_{00}$ respectively. In presenting the results of this work, we adhere to established convention and assume that the electron sector remains Lorentz invariant, as it has been for all modern cavity-based Michelson-Morley experiments.
\\
\\
Bounds for coefficients of the minimal SME are presented in Table~\ref{tab:results}. It is important to note that $\tilde{\kappa}_{\text{e}-}^{ZZ}$ is constrained solely by the amplitude of the cosine variation at twice the turntable frequency, 2$\omega_{\text{R}}$, which is dominated by known systematic noise processes (magnetic field). Therefore, the presented bound is obtained by taking the statistical mean of the time-dependent $C_{2}$ amplitudes from equation~\eqref{eq:freq}, with the error given by the standard deviation of the amplitudes. All coefficients are statistically insignificant; we report no evidence for violations of Lorentz symmetry in electrodynamics.
\\
\\
For the rotation symmetry breaking coefficients, $\tilde{\kappa}_{\text{e}-}$, our bounds are comparable to recent results from a trapped ion experiment~\cite{prutt15} that put constraints on four sets of linear combinations of electron and photon coefficients. In contrast, our reported bounds on the five $\tilde{\kappa}_{\text{e}-}$ coefficients were set under the assumption that Lorentz invariance of the electron is conserved (i.e. $c^{jk}$=0); ultimately this works derives bounds from nine linear combinations of photon and electron coefficients (see Supplementary Table 2), which in principle offers better opportunities for disentangling coefficients. Compared to bounds derived from previous modern Michelson-Morley cavity tests~\cite{Schiller2009,optical6}, where it was also assumed that the electron remained Lorentz invariant, we improve by up to a factor of $\sim$4.
\\
\\
The trapped ion experiment~\cite{prutt15} was not sensitive to the isotropic deviation, $\tilde{\kappa}_{\text{tr}}$, or the boost symmetry breaking coefficients, $\tilde{\kappa}_{\text{o}+}$. For $\tilde{\kappa}_{\text{o}+}$ we improve upon the current state-of-the-art~\cite{datatables,Schiller2009,optical6} by up to a factor of $\sim$5. We also improve upon our existing bounds~\cite{cav2} for the isotropic shift, $\tilde{\kappa}_{\text{tr}}$, by a factor of $\sim$20, as recent work~\cite{Mewes2012} demonstrated that double pass asymmetric resonators constructed out of a single material are only sensitive to higher order coefficients (d$>$4) of the SME.
\\
\\
Despite thoughtful advances in the implementation and design of alternative experiments, cavity-based Michelson-Morley tests such as the one presented in this work still remain the most powerful and comprehensive way to search for non-birefringent LIV effects in electrodynamics.
\\
\\
Scheduled upgrades to the experiment to improve the fundamental stability of the microwave oscillators~\cite{nand2013} will soon allow for even more sensitive tests. The installation of superior magnetic shielding around the oscillators will reduce the influence of systematic noise sources. The addition of a separate optical cavity system in the same cryogenic environment will also open up opportunities for exploring additional higher order coefficients~\cite{Mewes2012,parkerhigherorder} and matter sector coefficients~\cite{cav7}, allowing a more complete disentangling of electron and photon sector coefficients.
\\
\\
A modern Michelson-Morley experiment has been performed with two orthogonally aligned stable microwave oscillators. Using data of the beat note frequency between the two oscillators recorded over the course of a year we are able to constrain LIV-induced $\Delta\nu$/$\nu$ to be less than 10$^{-18}$, the most precise measurement ever made for electromagnetic cavity experiments. No violations of Lorentz symmetry were observed.
\\
\\\section{Methods}
\textbf{Experiment}\\\\
Two nominally identical cylinders (51~mm diameter, 30~mm height) of HEMEX grade ultra-pure sapphire (GT Advanced Technologies) were machined from the same boule. The crystals were cleaned in a solution of 70~$\%$ nitric acid containing several drops of hydrofluoric acid and then mounted in copper cavities and sealed in a stainless steel vacuum can, evacuated to sub-10$^{-6}$~mbar. Whispering gallery resonant mode WGE$_{16,0,0}$ was used for both cavities. This mode features a dominant radial electric field with 32 variations around the circumference of the sapphire crystal. The majority of the electromagnetic fields are contained within the sapphire dielectric, with minimal evanescent field leaking out to the copper walls of the cavity structure. The first cavity had a resonant frequency of 12.9688~GHz and a loaded quality factor of 10$^{9}$ at 4.4~K, while the second cavity had a resonant frequency of 12.9685~GHz and a loaded quality factor of 1.5$\times$10$^{9}$ at 4.4~K.
\\
\\
Small concentrations of impurities within the sapphire give rise to a temperature / frequency turning point~\cite{Locke:2008}; for the 347~kHz beat frequency between the two resonators this turning point occurred at 5.5~K. A temperature controller (model 340, Lake Shore Cryotronics, Inc.) was used in conjunction with a resistive heater and a carbon-glass temperature sensor to operate the resonators at the turning point. Custom microwave circuits and control electronics are used to create a loop oscillator out of each resonator. Pound locking is employed whereby modulation sidebands reflected back from the resonator are demodulated with a lock-in amplifier (model SR830, Standford Research Systems) and from this a correction signal is applied to a voltage controlled phase shifter to align the frequency of the oscillator with that of the resonator. Power incident on the cavity is monitored with a detector (model DT8016, Herotek, Inc.), the signal is compared against a user-defined set-point and a correction voltage is applied to a voltage controlled attenuator placed in situ with the loop oscillator. The oscillator comparison beat frequency was logged on a frequency counter (model 53142A, Agilent Technologies, Inc.) referenced to a 10~MHz rubidium standard.
\\
\\
The oscillators were rotated on a high-precision air-bearing rotation table (Kugler GmbH); an 18,000 point optical encoder was used to track the angular position of the table and maintain a constant rotation velocity. The table sat upon three aluminium legs, with each one in turn placed on a force sensor; these were used to align the centre of mass with the axis of rotation. A bi-axial high-gain tilt sensor (model 755, Applied Geomechanics) sits at the centre of the experiment. Variations in tilt were compensated for by heating or cooling two of the three aluminium legs. All three legs were heated above ambient temperature to improve performance of the tilt-control system.
\\
\\
\textbf{Data analysis}\\\\
Original time tags are converted into time in seconds since the Vernal Equinox prior to the start of data collection (March 20th 2012, 05:14 UTC+0). This format assists with calculating the relevant phase offsets required to analyse the data in the context of the SME~\cite{Kostelecky:2002}. The rotation turntable features an optical encoder with 18,000 points and a trigger mark to indicate that a full rotation has occurred; the data is scanned and incomplete rotations are discarded. The rotation points are converted into a modular angle value in radians. The data is broken up into subsets containing 10 full turntable rotations (corresponding to $\sim$1000~seconds) and an ordinary least squares regression is used to fit the subset to the following model:
\begin{align}
\nu_{\text{beat}}=A+Bt+C_{\omega_{\text{R}}}\cos{\left(\omega_{\text{R}}t+\phi_{\text{R}}\right)}+S_{\omega_{\text{R}}}\sin{\left(\omega_{\text{R}}t+\phi_{\text{R}}\right)} \nonumber \\
+C_{2\omega_{\text{R}}}\cos{\left(2\omega_{\text{R}}t+2\phi_{\text{R}}\right)}+S_{2\omega_{\text{R}}}\sin{\left(2\omega_{\text{R}}t+2\phi_{\text{R}}\right)}. \label{eq:firstdls}
\end{align}
The value of $\omega_{\text{R}}t$ comes straight from the modular angular position of the turntable recorded in the data. The phase offset, $\phi_{\text{R}}$, is the angular difference between the start of data collection and the alignment of the crystal-axis of the top cavity with geographical East. The start of data collection is triggered at the same point for each experimental run. The phase offset is once again only needed to aid with the reporting of coefficients in the SME framework. The amplitudes of equation~\eqref{eq:firstdls} and the mean time of each subset are stored to file, producing 19,597 entries. A histogram of the magnitude of errors for the fits to $C_{2\omega_{\text{R}}}$ and $S_{2\omega_{\text{R}}}$ is produced (Supplementary Figure 9), all subsets with an error further than 3$\sigma$ from the mean are discarded, resulting in 299 entries being removed (1.5$\%$ of the full dataset). These points correspond to data with excessive additional noise present that does not fit our model or expected signals and would otherwise corrupt the quality of the subsequent analysis.
\\
\\
The demodulated dataset is then broken up into subsets containing 100 entries each ($\sim$1.2 days) and fit to the following model via a weighted least squares regression, using the square of the standard errors of the fits from equation~\eqref{eq:firstdls} as the weights.
\begin{align}
C_{2\omega_{\text{R}}}\left(t\right)=C_{0}+CC_{\omega_{\oplus}}\cos{\left(\omega_{\oplus}t+\phi_{\oplus}\right)}+CS_{\omega_{\oplus}}\sin{\left(\omega_{\oplus}t+\phi_{\oplus}\right)} \nonumber \\
+CC_{2\omega_{\oplus}}\cos{\left(2\omega_{\oplus}t+2\phi_{\oplus}\right)}+CS_{2\omega_{\oplus}}\sin{\left(2\omega_{\oplus}t+2\phi_{\oplus}\right)} \label{eq:dlsC2} \\
S_{2\omega_{\text{R}}}\left(t\right)=S_{0}+SC_{\omega_{\oplus}}\cos{\left(\omega_{\oplus}t+\phi_{\oplus}\right)}+SS_{\omega_{\oplus}}\sin{\left(\omega_{\oplus}t+\phi_{\oplus}\right)} \nonumber \\
+SC_{2\omega_{\oplus}}\cos{\left(2\omega_{\oplus}t+2\phi_{\oplus}\right)}+SS_{2\omega_{\oplus}}\sin{\left(2\omega_{\oplus}t+2\phi_{\oplus}\right)} \label{eq:dlsS2}
\end{align}
The phase offset $\phi_{\oplus}$ is the difference between the 2012 Vernal Equinox and the alignment of geographical East with the Y axis of the Sun Centred frame used for determination of the SME coefficients~\cite{Kostelecky:2002}. The value of $\omega_{\oplus}$ used was 7.3$\times$10$^{-5}$ rad/s and $t$ is the relevant mean time calculated during the previous demodulation (equation~\ref{eq:firstdls}). Once again the amplitudes, standard errors and mean time for each subset is stored to file. The fitted amplitudes and standard errors from equations~\eqref{eq:dlsC2} and~\eqref{eq:dlsS2} are shown in Supplementary Figures~1~--~8. The 8 amplitudes from equations~\eqref{eq:dlsC2} and~\eqref{eq:dlsS2} representing the first two harmonics of daily variations are used to bound the overall sensitivity of the experiment to LIVs. We take the standard error-weighted average of all the amplitudes, which is equivalent to weighting by the variance. Noting the distribution of the histogram of all the amplitudes (Supplementary Figure 10), whereby 95$\%$ of the points lie within 2 standard deviations of the mean, we multiply the associated standard error by 2 to determine the 95$\%$ confidence interval for our bound, $\Delta\nu$/$\nu$~$\leq$9.2$\pm$10.7$\times$10$^{-19}$.
\\
\\
The final stage of the data analysis is used to set bounds on coefficients of the SME. Each amplitude and standard error from equations~\eqref{eq:dlsC2} and~\eqref{eq:dlsS2} is used to perform a weighted least squares regression fit to an offset and variations at harmonics of Earth's orbital frequency, $\Omega_{\oplus}$ and 2$\Omega_{\oplus}$. Supplementary Table~2 summarizes the relevant amplitudes, their sensitivity to different coefficients of the SME and the corresponding numerical weights.
\\
\\
For SME coefficients where more than one bound is derived from Supplementary Table~2 we report the error-weighted average of all contributions. The coefficient $\tilde{\kappa}_{\text{e}-}^{ZZ}$ is only accessible via the second harmonic of turntable rotation (amplitude $C_{2\omega_{\text{R}}}$ in equation~\eqref{eq:firstdls}, or offset $C_{0}$ in equation~\eqref{eq:dlsC2}), where systematic noise sources are present. The bound for this coefficient is set by taking the statistical mean and standard deviation of all the values obtained for $C_{0}$, noting that the bound should be consistent with a null result for systematic noise with a random phase, while a substantial Lorentz violating signal with constant phase would be statistically significant.
\\
\\
\textbf{Acknowledgements} \\
This work was supported by Australian Research Council Discovery Project DP130100205 and by the Go8-DAAD Australia-Germany Joint Research Co-operation Scheme.

\clearpage

\section*{Supplementary Figures}
\begin{figure}[h!]
\centering
\includegraphics[width=0.75\columnwidth]{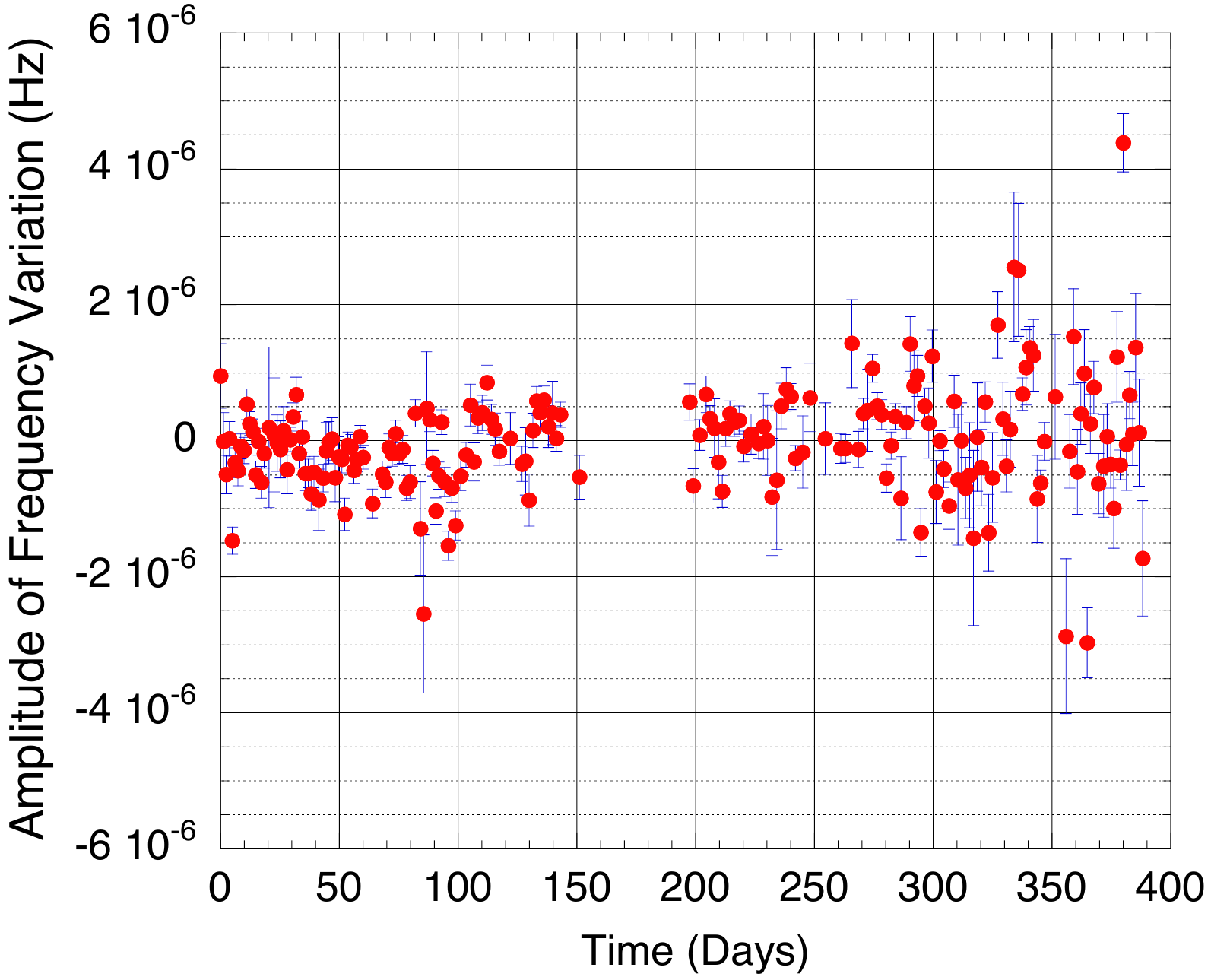}
\caption{\textbf{Subset fits to sidereal amplitude $CC_{\omega_{\oplus}}$} from equation~(3) obtained as discussed in the main text. Statistical 1$\sigma$ error bars are shown in blue.}
\label{fig:CC1w}
\end{figure}
\begin{figure}[h!]
\centering
\includegraphics[width=0.75\columnwidth]{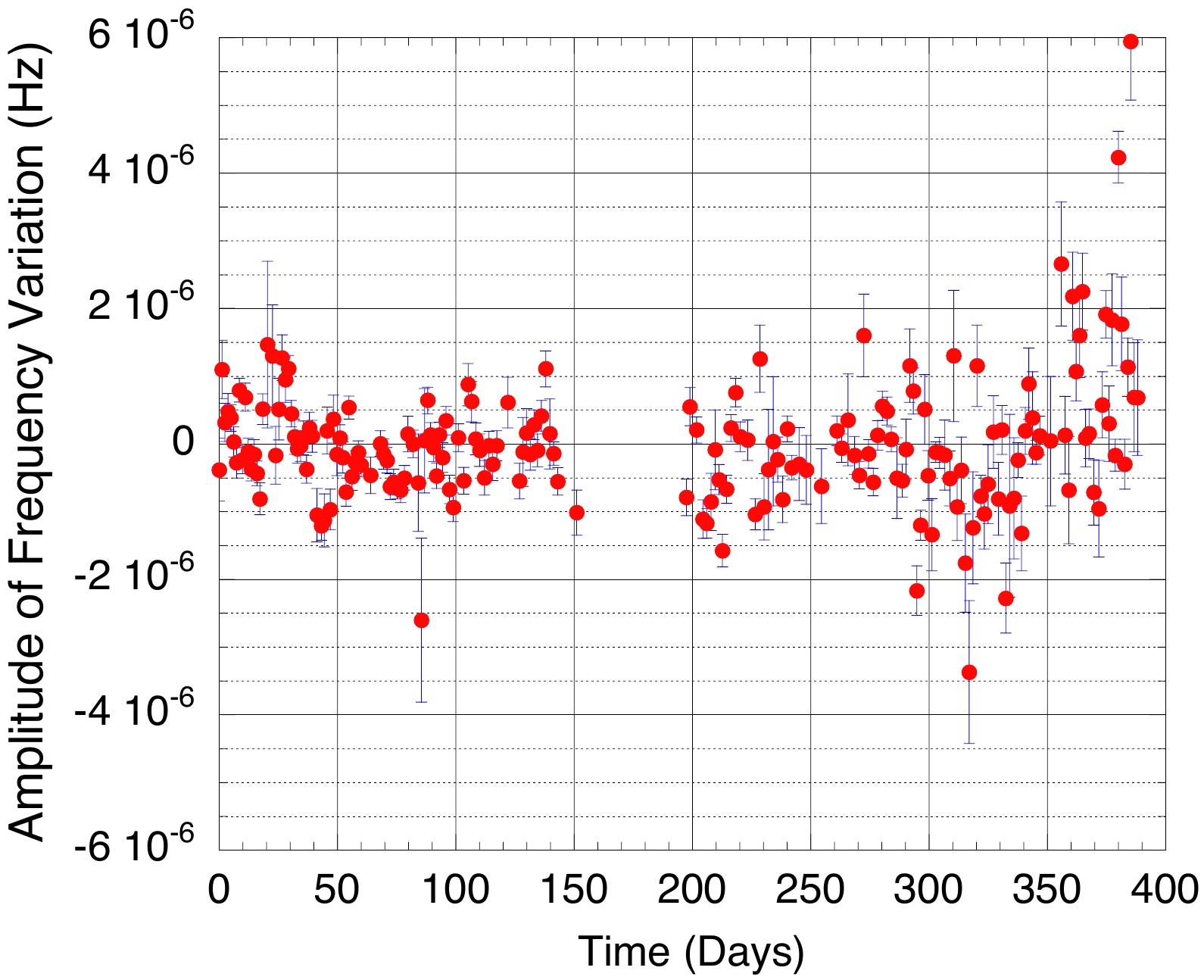}
\caption{\textbf{Subset fits to sidereal amplitude $CS_{\omega_{\oplus}}$} from equation~(3) obtained as discussed in the main text. Statistical 1$\sigma$ error bars are shown in blue.}
\label{fig:CS1w}
\end{figure}
\begin{figure}[h!]
\centering
\includegraphics[width=0.74\columnwidth]{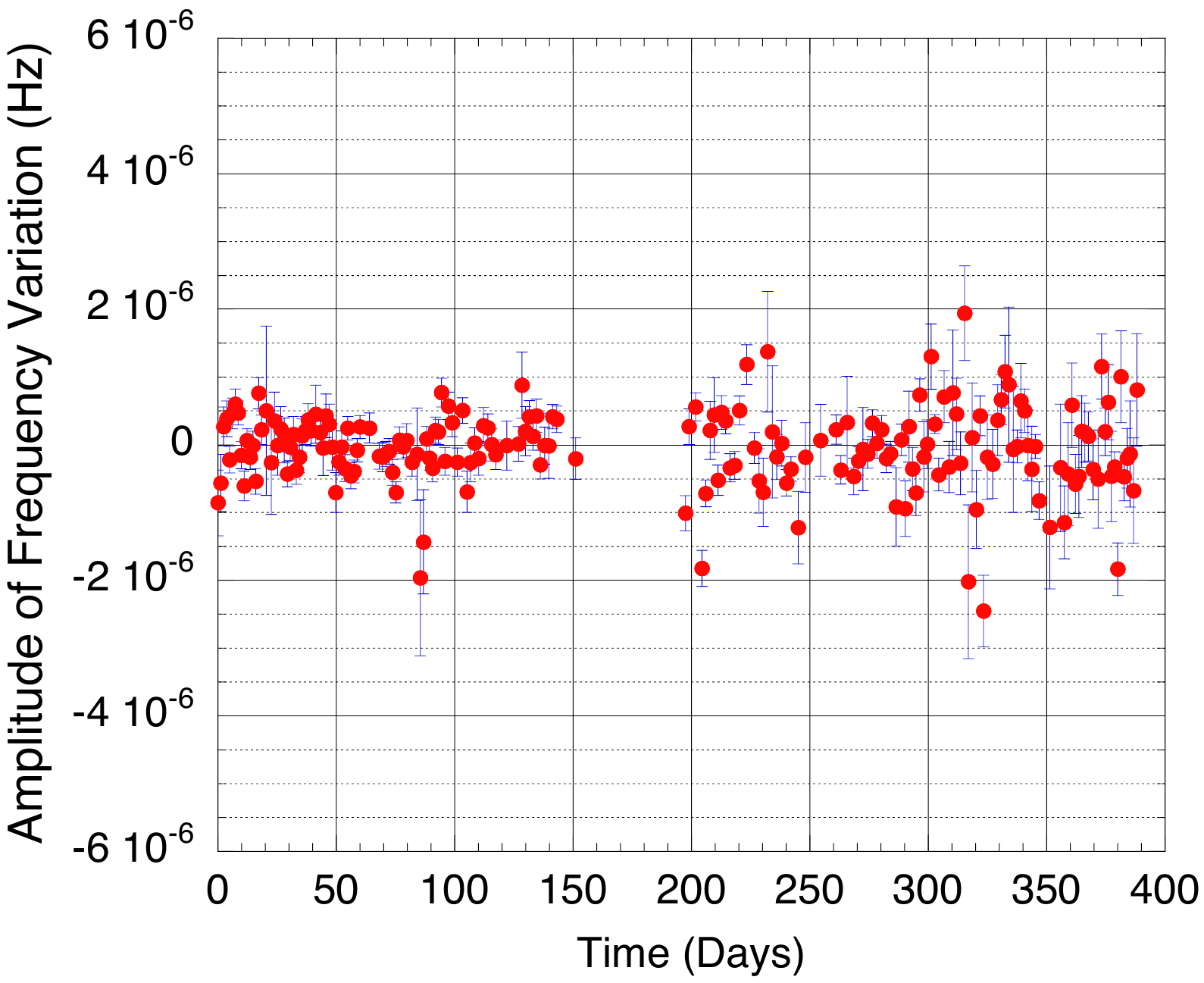}
\caption{\textbf{Subset fits to sidereal amplitude $CC_{2\omega_{\oplus}}$} from equation~(3) obtained as discussed in the main text. Statistical 1$\sigma$ error bars are shown in blue.}
\label{fig:CC2w}
\end{figure}
\begin{figure}[h!]
\centering
\includegraphics[width=0.75\columnwidth]{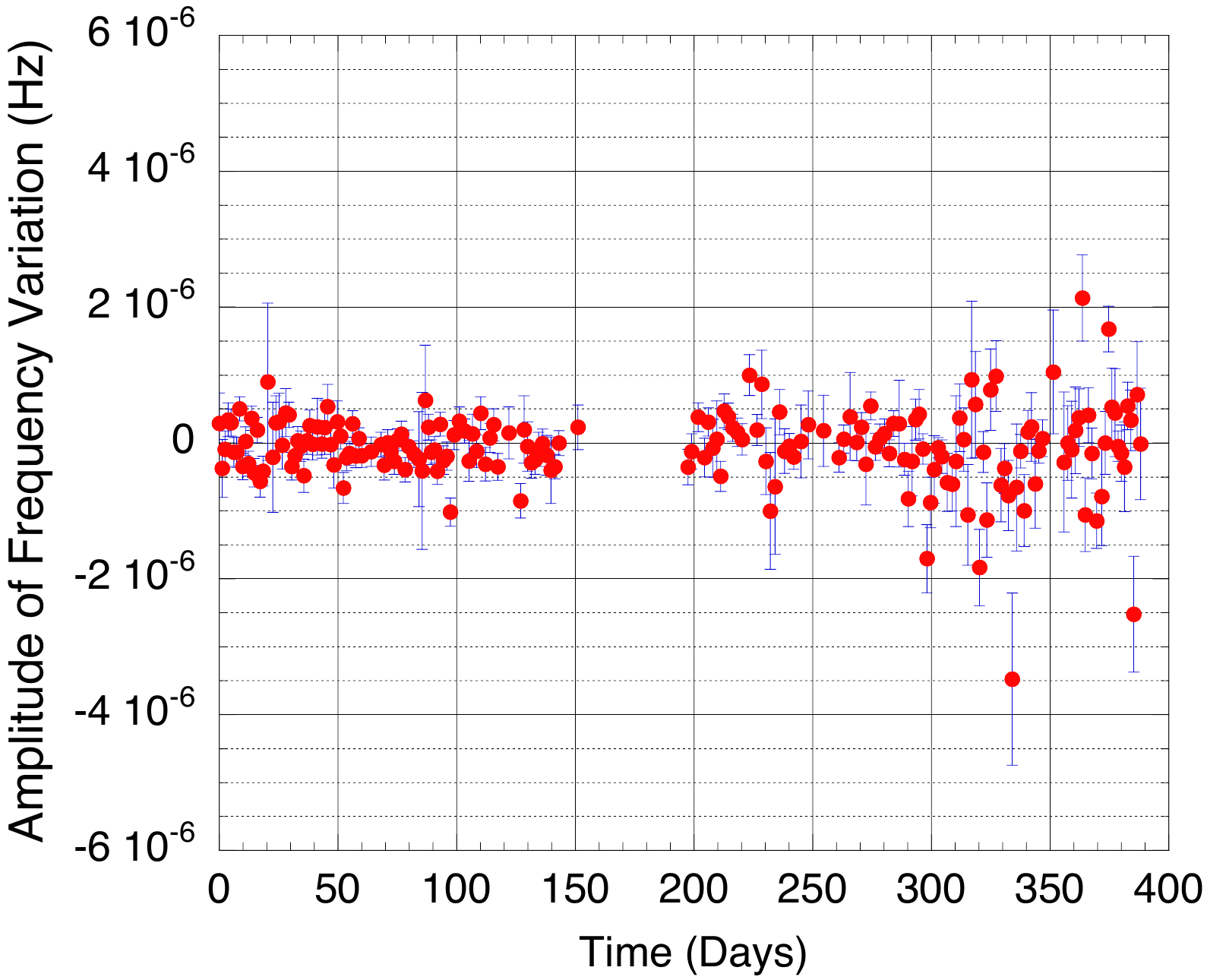}
\caption{\textbf{Subset fits to sidereal amplitude $CS_{2\omega_{\oplus}}$} from equation~(3) obtained as discussed in the main text. Statistical 1$\sigma$ error bars are shown in blue.}
\label{fig:CS2w}
\end{figure}
\begin{figure}[h!]
\centering
\includegraphics[width=0.75\columnwidth]{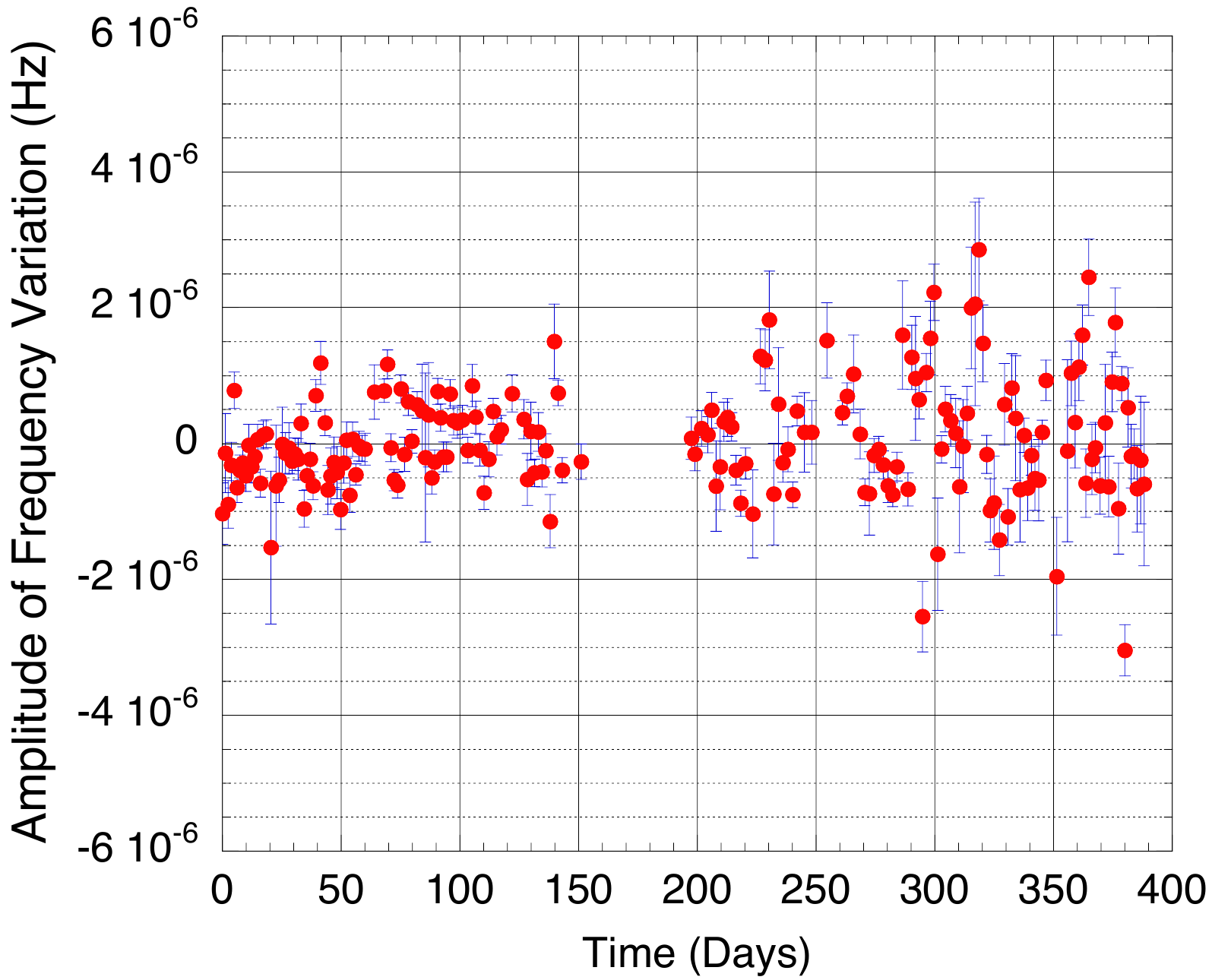}
\caption{\textbf{Subset fits to sidereal amplitude $SC_{\omega_{\oplus}}$} from equation~(4) obtained as discussed in the main text. Statistical 1$\sigma$ error bars are shown in blue.}
\label{fig:SC1w}
\end{figure}
\begin{figure}[h!]
\centering
\includegraphics[width=0.75\columnwidth]{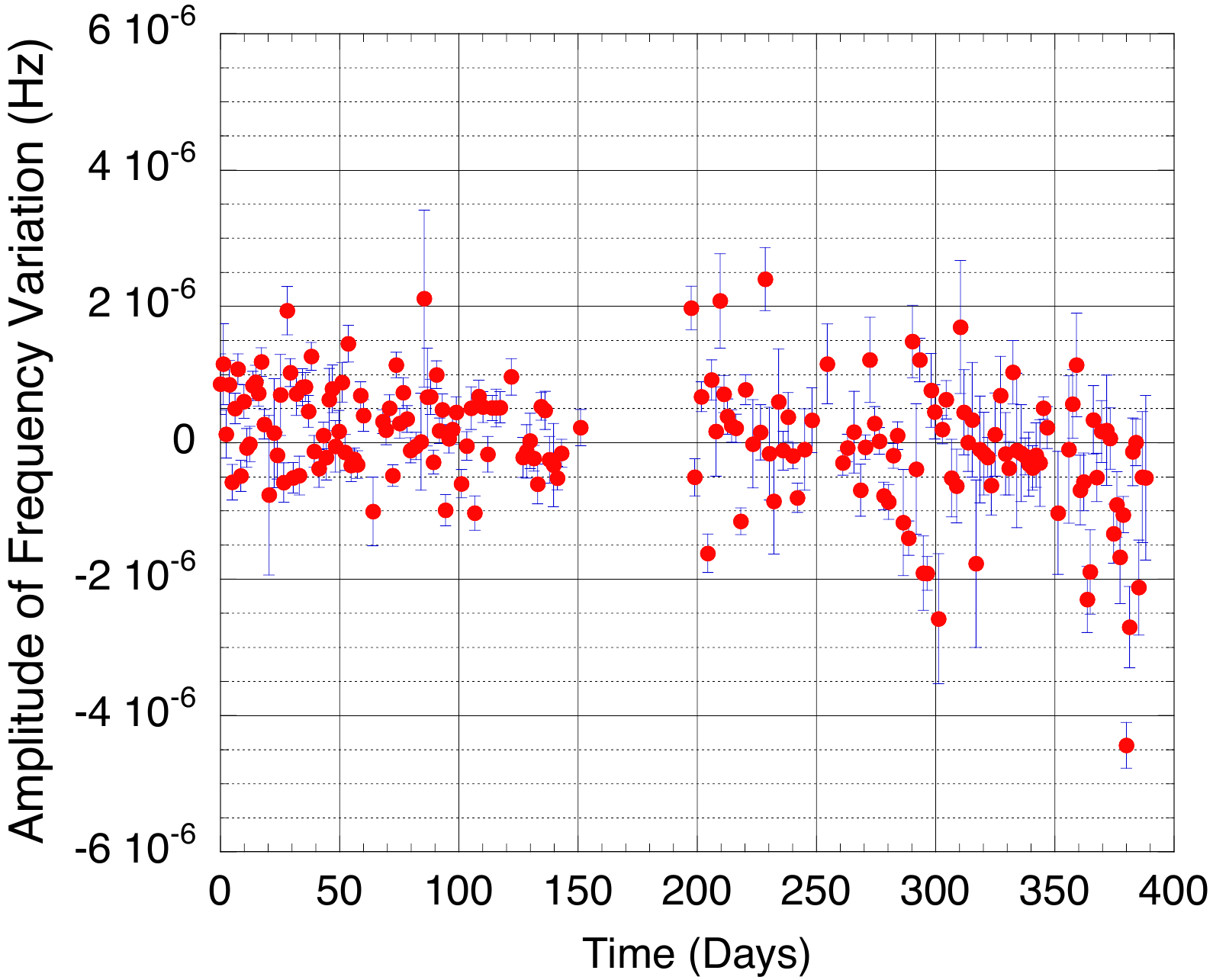}
\caption{\textbf{Subset fits to sidereal amplitude $SS_{\omega_{\oplus}}$} from equation~(4) obtained as discussed in the main text. Statistical 1$\sigma$ error bars are shown in blue.}
\label{fig:SS1w}
\end{figure}
\begin{figure}[h!]
\centering
\includegraphics[width=0.74\columnwidth]{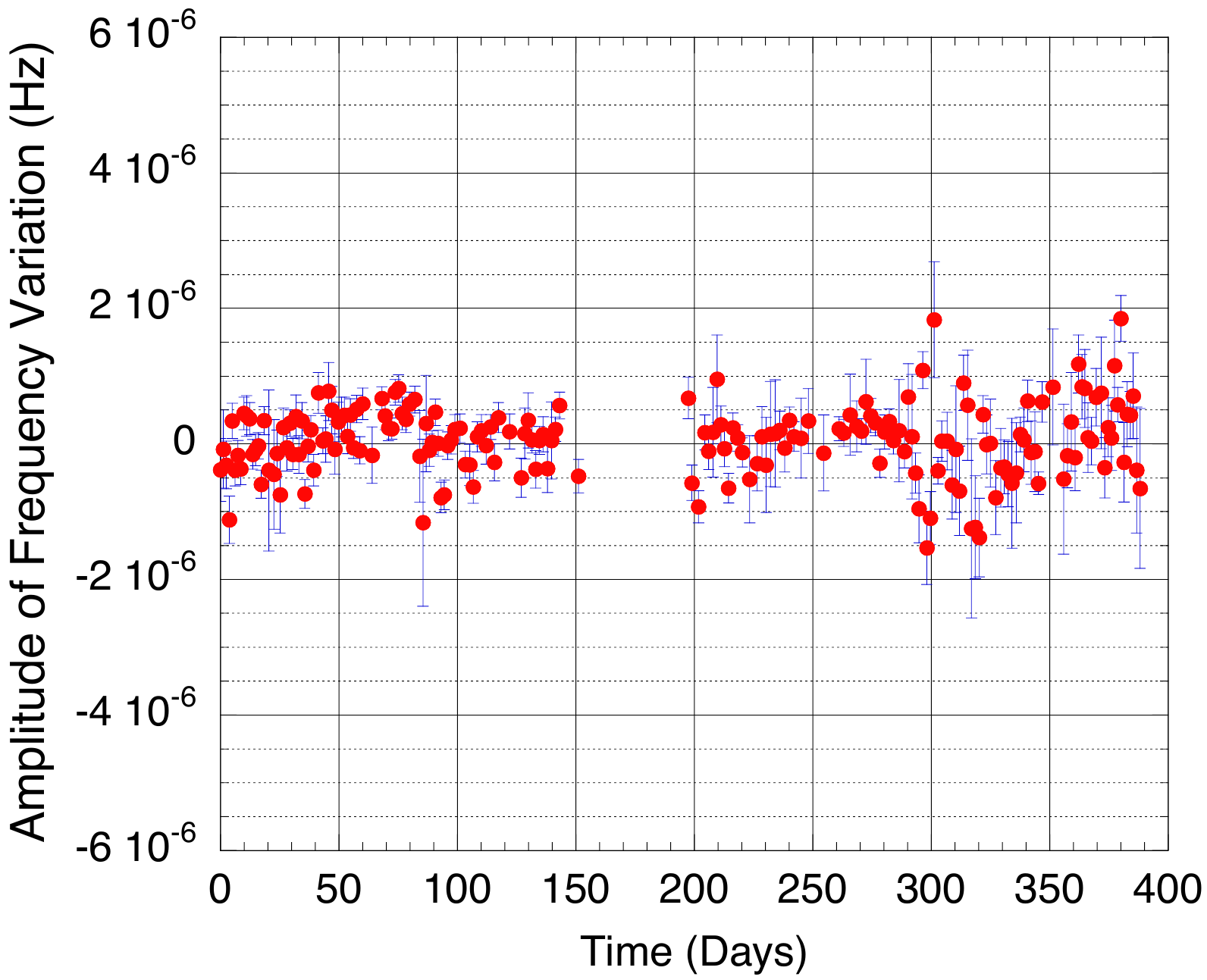}
\caption{\textbf{Subset fits to sidereal amplitude $SC_{2\omega_{\oplus}}$} from equation~(4) obtained as discussed in the main text. Statistical 1$\sigma$ error bars are shown in blue.}
\label{fig:SC2w}
\end{figure}
\begin{figure}[h!]
\centering
\includegraphics[width=0.72\columnwidth]{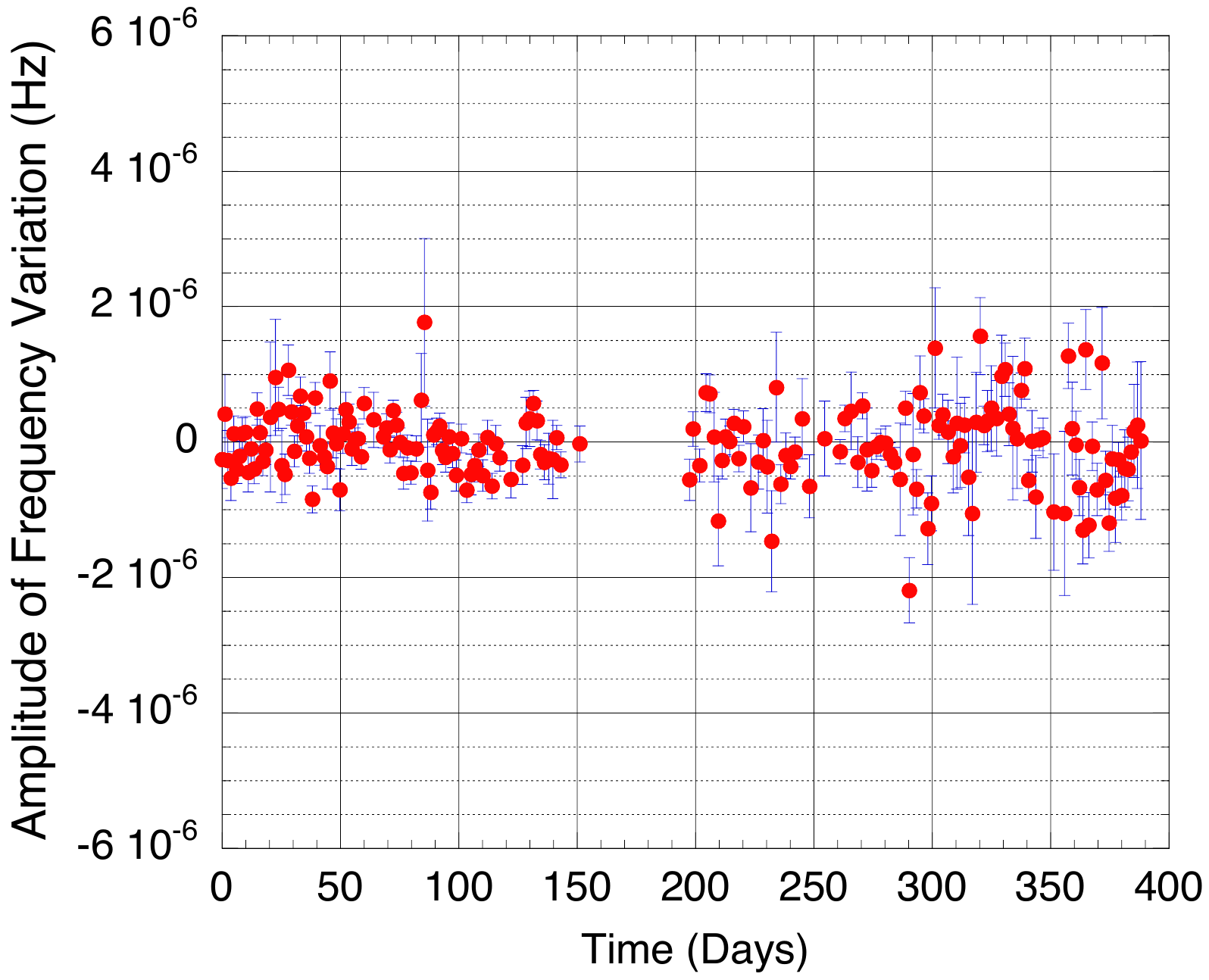}
\caption{\textbf{Subset fits to sidereal amplitude $SS_{2\omega_{\oplus}}$} from equation~(4) obtained as discussed in the main text. Statistical 1$\sigma$ error bars are shown in blue.}
\label{fig:SS2w}
\end{figure}
\begin{figure}[h!]
\centering
\includegraphics[width=0.72\columnwidth]{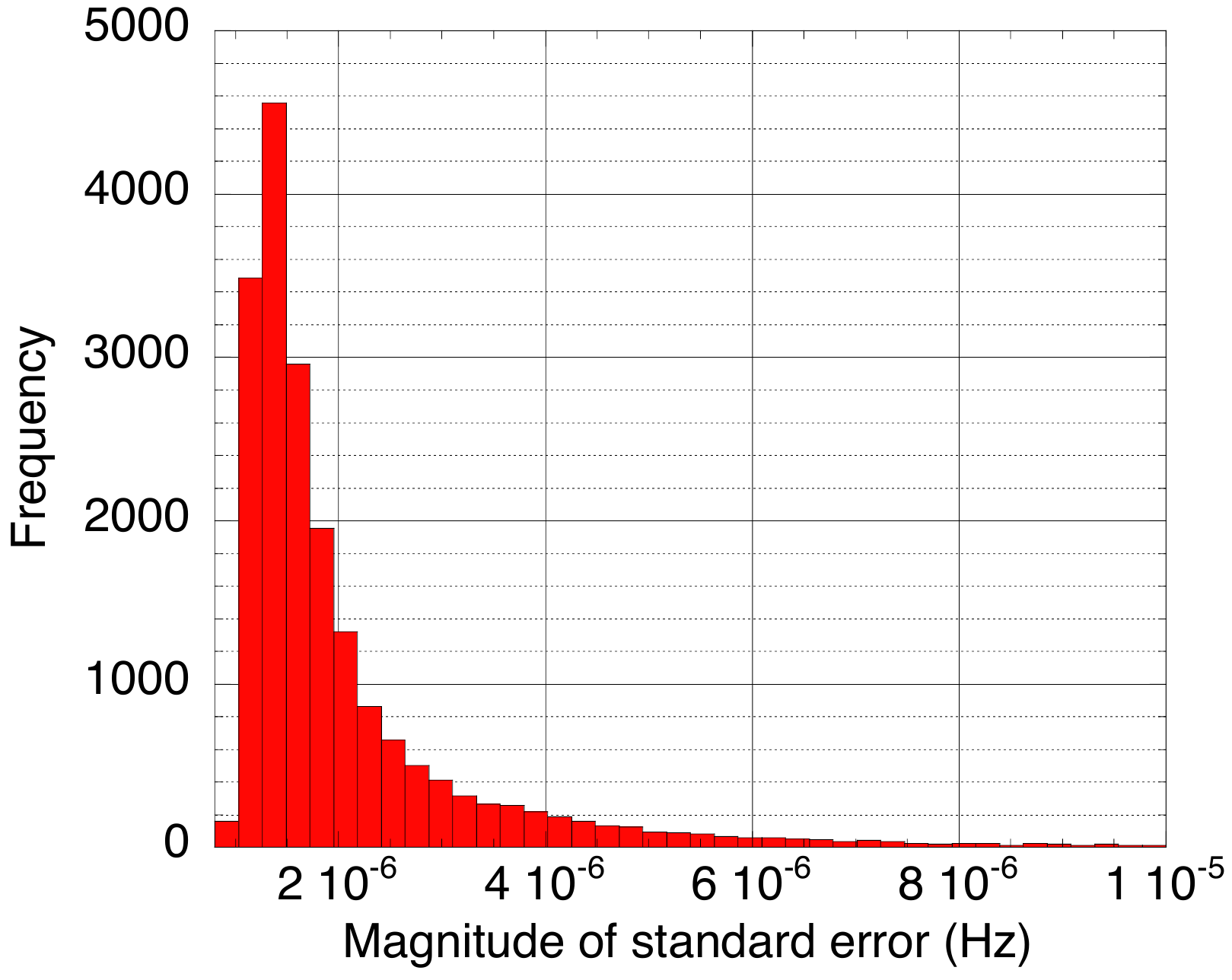}
\caption{\textbf{Statistical histogram of the magnitude of the standard errors} for fits to the amplitudes of the 2$\omega_{R}$ components of equation~(2) in the main text. Values are from fits to subsets of data $\sim$1000 seconds (10 rotations) long.}
\label{fig:dlshist}
\end{figure}
\begin{figure}[h!]
\centering
\includegraphics[width=0.75\columnwidth]{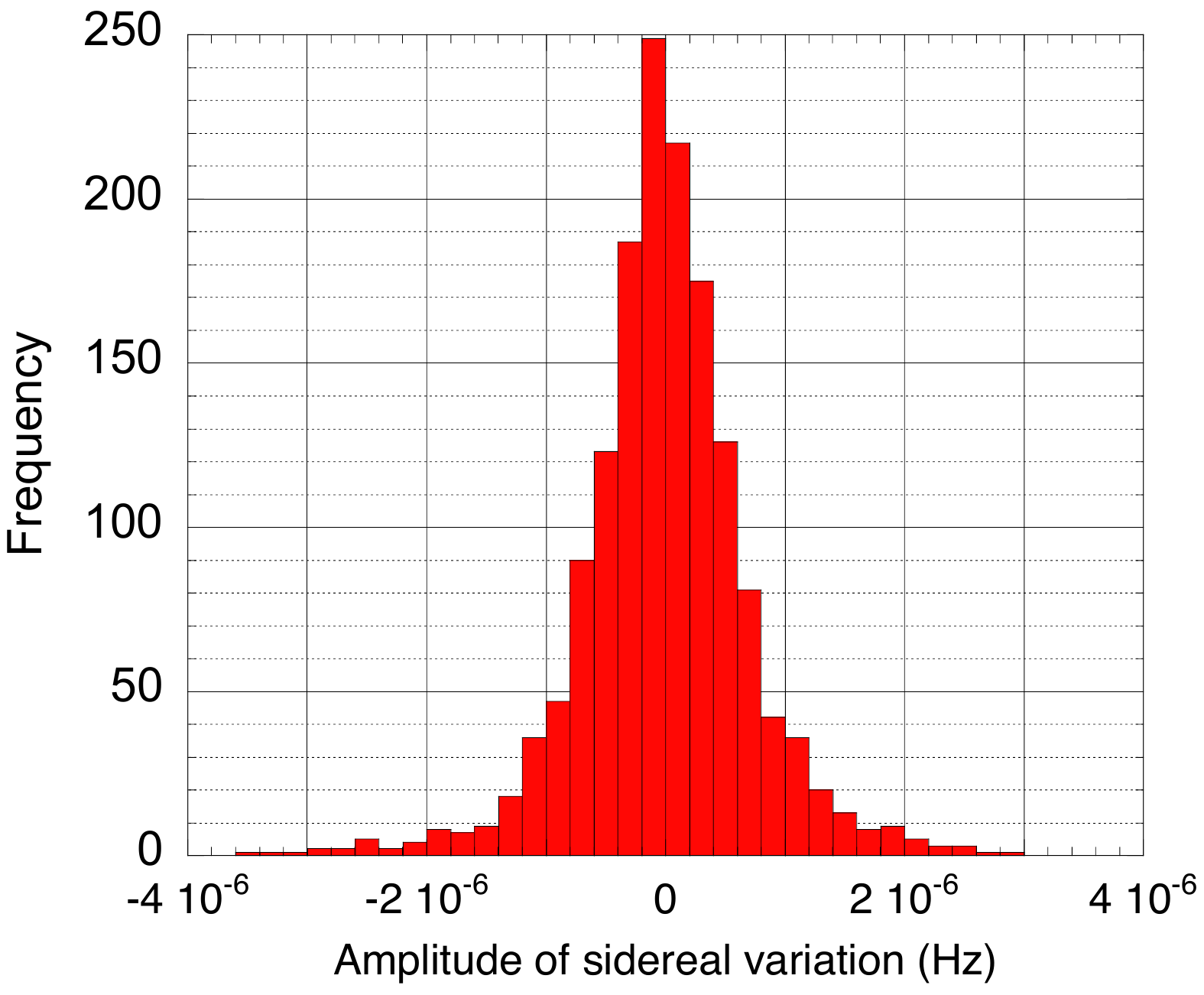}
\caption{\textbf{Statistical histogram of amplitudes for $\omega_{\oplus}$ and 2$\omega_{\oplus}$ variations}, given by the 8 time-varying components of equations~(3) and~(4) in the main text. Values are from fits to subsets of data $\sim$1.2 days long.}
\label{fig:sidhist}
\end{figure}
\section*{Supplementary Tables}
\begin{table}[h!]
\caption{\textbf{Historical overview of Michelson-Morley style experiments}, constraining Lorentz invariance in electrodynamics as presented in Figure~(1) of the main text.}
\centering
\begin{tabular}{c|c|c|c}
\hline\hline
Year & Type & $\Delta c$/$c$ or $\Delta\nu$/$\nu$ & Reference \\
\hline
1881 & Interferometer & 8.3E-9 & \cite{1881} \\
1887 & Interferometer & 9.1E-10 & \cite{1887} \\
1904 & Interferometer & 2.3E-10 & \cite{1904} \\
1924 & Interferometer & 1.2E-9 & \cite{1924} \\
1925 & Interferometer & 1.4E-9 & \cite{1926a} \\
1926 & Interferometer & 5E-10 & \cite{1926b} \\
1927 & Interferometer & 1E-10 & \cite{1927} \\
1930 & Interferometer & 4.8E-11 & \cite{1930} \\
1955 & Cavities & 1E-10 & \cite{1955} \\
1964 & Interferometer & 1E-11 & \cite{1964} \\
1969 & Interferometer & 8.1E-10 & \cite{1969} \\
1979 & Cavities & 4E-15 & \cite{1979} \\
2003 & Cavities & 3.4E-14 & \cite{2003a} \\
2003 & Cavities & 1E-15 & \cite{2003b} \\
2003 & Cavities & 4.3E-15 & \cite{2003c} \\
2004 & Cavities & 1.1E-15 & \cite{2004} \\
2005 & Cavities & 2.6E-16 & \cite{2005a} \\
2005 & Cavities & 5E-17 & \cite{2005b} \\
2006 & Cavities & 8E-17 & \cite{2006} \\
2009 & Cavities & 1E-17 & \cite{2009a} \\
2009 & Cavities & 1E-17 & \cite{2009b} \\
2014 & Cavities & 1E-18 & This work \\
\hline\hline
\end{tabular}
\label{tab:histiso}
\end{table}
\begin{table}[h!]
\caption[Results]{\textbf{Amplitudes of Cosine / Sine frequency components of interest and their sensitivities and numerical weights}, calculating using the following values: $F_{1,2}$ describe combinations of electric and magnetic filling factors in the cavities ($\sim$0.46,0.5), $\chi$ is the co-latitude of the experiment from the north pole ($\sim38^{\circ}$), $\eta$ is the angle between Earth's orbital and equatorial planes ($\sim$23.4$^{\circ}$) and $\beta_{\oplus}$, Earth's orbital velocity suppressed by the canonical value for the speed of light in vacuum (9.9E-5). These sensitivities arise from orientation and design of the experiment and the frame transformations required to express bounds in the sun-centred frame of choice. Frame transformations and determination of sensitivities is discussed at length in the literature.}
\centering
\begin{tabular}{c|c|c}
\hline\hline
Amplitude & Sensitivity & Numerical Weight \\
\hline
$S_{0}$ & - & - \\
\hline
$SS_{\omega_{\oplus}}^{0}$ & 4$F_{1}\sin{\left(\chi \right)}\tilde{\kappa}_{\text{e}-}^{YZ}$ & 1.1  \\
$SS_{\omega_{\oplus}}C_{\Omega_{\oplus}}$ & -4$F_{2}\beta_{\oplus}\sin{\left(\chi\right)}\left(\cos{\left(\eta\right)}\tilde{\kappa}_{\text{o}+}^{XY}-\sin{\left(\eta\right)}\tilde{\kappa}_{\text{o}+}^{XZ}\right)$ & 4.9E-5 $\tilde{\kappa}_{\text{o}+}^{XZ}$-1.1E-4 $\tilde{\kappa}_{\text{o}+}^{XY}$ \\
$SS_{\omega_{\oplus}}C_{2\Omega_{\oplus}}$ & -$F_{1}\beta_{\oplus}^{2}\sin{\left(2\eta\right)}\sin{\left(\chi\right)}\tilde{\kappa}_{\text{tr}}$ & -2E-9 \\
\hline
$SC_{\omega_{\oplus}}^{0}$ & -4$F_{1}\sin{\left(\chi\right)}\tilde{\kappa}_{\text{e}-}^{XZ}$ & -1.1 \\
$SC_{\omega_{\oplus}}S_{\Omega_{\oplus}}$ & 4$F_{2}\beta_{\oplus}\sin{\left(\chi\right)}\tilde{\kappa}_{\text{o}+}^{XY}$ & 1.2E-4 \\
$SC_{\omega_{\oplus}}C_{\Omega_{\oplus}}$ & 4$F_{2}\beta_{\oplus}\sin{\left(\eta\right)}\sin{\left(\chi\right)}\tilde{\kappa}_{\text{o}+}^{YZ}$ & 4.8E-5 \\
$SC_{\omega_{\oplus}}S_{2\Omega_{\oplus}}$ & 2$F_{1}\beta_{\oplus}^{2}\sin{\left(\eta\right)}\sin{\left(\chi\right)}\tilde{\kappa}_{\text{tr}}$ & 2.2E-9 \\
\hline
$SS_{2\omega_{\oplus}}^{0}$ & -4$F_{1}\cos{\left(\chi\right)}\tilde{\kappa}_{\text{e}-}^{XY}$ & -1.5 \\
$SS_{2\omega_{\oplus}}S_{\Omega_{\oplus}}$ & -4$F_{2}\beta_{\oplus}\cos{\left(\chi\right)}\tilde{\kappa}_{\text{o}+}^{XZ}$  & -1.6E-4 \\
$SS_{2\omega_{\oplus}}C_{\Omega_{\oplus}}$ & 4$F_{2}\beta_{\oplus}\cos{\left(\eta\right)}\cos{\left(\chi\right)}\tilde{\kappa}_{\text{o}+}^{YZ}$  & 1.5E-4 \\
$SS_{2\omega_{\oplus}}S_{2\Omega_{\oplus}}$ & 2$F_{1}\beta_{\oplus}^{2}\cos{\left(\eta\right)}\cos{\left(\chi\right)}\tilde{\kappa}_{\text{tr}}$ & 6.6E-9 \\
\hline
$SC_{2\omega_{\oplus}}^{0}$ & 2$F_{1}\cos{\left(\chi\right)}\left(\tilde{\kappa}_{\text{e}-}^{XX}-\tilde{\kappa}_{\text{e}-}^{YY}\right)$ & 7.2E-1 \\
$SC_{2\omega_{\oplus}}C_{\Omega_{\oplus}}$ & -4$F_{2}\beta_{\oplus}\cos{\left(\eta\right)}\cos{\left(\chi\right)}\tilde{\kappa}_{\text{o}+}^{XZ}$ & -1.5E-4 \\
$SC_{2\omega_{\oplus}}S_{\Omega_{\oplus}}$ & -4$F_{2}\beta_{\oplus}\cos{\left(\chi\right)}\tilde{\kappa}_{\text{o}+}^{YZ}$ & -1.6E-4 \\
$SC_{2\omega_{\oplus}}C_{2\Omega_{\oplus}}$ & 0.5$F_{1}\beta_{\oplus}^{2}\left(3+\cos{\left(2\eta\right)}\right)\cos{\left(\chi\right)}\tilde{\kappa}_{\text{tr}}$ & 6.6E-9 \\
\hline
$C_{0}$ & -3$F_{1}\sin{\left(\chi\right)}^{2}\tilde{\kappa}_{\text{e}-}^{ZZ}$ & -5.1E-1 \\
\hline
$CS_{\omega_{\oplus}}^{0}$ & 4$F_{1}\sin{\left(\chi \right)}\tilde{\kappa}_{\text{e}-}^{YZ}$ & 8.8E-1  \\
$CS_{\omega_{\oplus}}C_{\Omega_{\oplus}}$ & 2$F_{2}\beta_{\oplus}\sin{\left(2\chi\right)}\left(\sin{\left(\eta\right)}\tilde{\kappa}_{\text{o}+}^{XZ}-\cos{\left(\eta\right)}\tilde{\kappa}_{\text{o}+}^{XY}\right)$ & 3.9E-5 $\tilde{\kappa}_{\text{o}+}^{XZ}$-8.9E-5 $\tilde{\kappa}_{\text{o}+}^{XY}$ \\
$CS_{\omega_{\oplus}}C_{2\Omega_{\oplus}}$ & 2$F_{1}\beta_{\oplus}^{2}\sin{\left(\eta\right)}\cos{\left(\eta\right)}\sin{\left(\chi\right)}\cos{\left(\chi\right)}\tilde{\kappa}_{\text{tr}}$ & 1.6E-9 \\
\hline
$CC_{\omega_{\oplus}}^{0}$ & 4$F_{1}\sin{\left(\chi \right)}\tilde{\kappa}_{\text{e}-}^{XZ}$ & 8.8E-1  \\
$CC_{\omega_{\oplus}}S_{\Omega_{\oplus}}$ & 2$F_{2}\beta_{\oplus}\sin{\left(2\chi \right)}\tilde{\kappa}_{\text{o}+}^{XY}$ & 9.7E-5 \\
$CC_{\omega_{\oplus}}C_{\Omega_{\oplus}}$ & -2$F_{2}\beta_{\oplus}\sin{\left(2\chi \right)}\sin{\left(\eta\right)}\tilde{\kappa}_{\text{o}+}^{YZ}$ & -3.8E-5 \\
$CC_{\omega_{\oplus}}S_{2\Omega_{\oplus}}$ & -$F_{1}\beta_{\oplus}^{2}\sin{\left(\eta\right)}\sin{\left(2\chi\right)}\tilde{\kappa}_{\text{tr}}$ & -1.7E-9 \\
\hline
$CS_{2\omega_{\oplus}}^{0}$ & -$F_{1}\left(3+\cos{\left(2\chi \right)}\right)\tilde{\kappa}_{\text{e}-}^{XY}$ & -1.5 \\
$CS_{2\omega_{\oplus}}C_{\Omega_{\oplus}}$ & $F_{2}\beta_{\oplus}\left(3+\cos{\left(2\chi\right)}\right)\cos{\left(\eta\right)}\tilde{\kappa}_{\text{o}+}^{YZ}$  & 1.5E-4 \\
\hline
$CC_{2\omega_{\oplus}}^{0}$ & -0.5$F_{1}\left(3+\cos{\left(2\chi \right)}\right)\left(\tilde{\kappa}_{\text{e}-}^{XX}-\tilde{\kappa}_{\text{e}-}^{YY}\right)$ & -7.4E-1 \\
$CC_{2\omega_{\oplus}}C_{\Omega_{\oplus}}$ & $F_{2}\beta_{\oplus}\left(3+\cos{\left(2\chi\right)}\right)\cos{\left(\eta\right)}\tilde{\kappa}_{\text{o}+}^{XZ}$ & 1.5E-4 \\ 
$CC_{2\omega_{\oplus}}S_{\Omega_{\oplus}}$ & -$F_{2}\beta_{\oplus}\left(3+\cos{\left(2\chi\right)}\right)\tilde{\kappa}_{\text{o}+}^{YZ}$ & -1.6E-4 \\ 
$CC_{2\omega_{\oplus}}C_{2\Omega_{\oplus}}$ & 0.13$F_{1}\beta_{\oplus}^{2}\left(3+\cos{\left(2\eta\right)}\right)\left(3+\cos{\left(2\chi\right)}\right)\tilde{\kappa}_{\text{tr}}$ & 6.8E-9 \\ 
\hline\hline
\end{tabular}
\label{tab:SMEweights}
\end{table}


\clearpage
\begin{thebibliography}{10}

\bibitem{mm1}
A.~A. Michelson and E.~W. Morley, ``On the relative motion of the earth and the
  luminiferous ether,'' {\em Am. J. Sci.}, vol.~34, pp.~333--345, Nov 1887.

\bibitem{bluhm2006}
R.~Bluhm, ``Overview of the standard model extension: Implications and
  phenomenology of lorentz violation,'' in {\em Special Relativity} (J.~Ehlers
  and C.~Lämmerzahl, eds.), vol.~702 of {\em Lecture Notes in Physics},
  pp.~191--226, Springer Berlin Heidelberg, 2006.

\bibitem{liberati2012}
S.~Liberati and D.~Mattingly, ``Lorentz breaking effective field theory models
  for matter and gravity: theory and observational constraints,'' in {\em
  Proceedings of the 2009 SIGRAV Graduate School in Contemporary Relativity and
  Gravitionational Physics}, 2009.

\bibitem{tasson2014}
J.~D. Tasson, ``What do we know about lorentz invariance?,'' {\em Reports on
  Progress in Physics}, vol.~77, no.~6, p.~062901, 2014.

\bibitem{weinberg2009}
S.~Weinberg, ``Effective field theory, past and future,'' in {\em Proceedings
  of the sixth International Workshop on Chiral Dynamics}, no.~UTTG-09-09,
  Proceedings of Science, 2009.

\bibitem{weinberg1967}
S.~Weinberg, ``A model of leptons,'' {\em Phys. Rev. Lett.}, vol.~19,
  pp.~1264--1266, Nov 1967.

\bibitem{ks89}
V.~A. Kosteleck\'y and S.~Samuel, ``Spontaneous breaking of lorentz symmetry in
  string theory,'' {\em Phys. Rev. D}, vol.~39, pp.~683--685, Jan 1989.

\bibitem{myers2003}
R.~C. Myers and M.~Pospelov, ``Ultraviolet modifications of dispersion
  relations in effective field theory,'' {\em Phys. Rev. Lett.}, vol.~90,
  p.~211601, May 2003.

\bibitem{petr2009}
P.~Ho\ifmmode~\check{r}\else \v{r}\fi{}ava, ``Quantum gravity at a lifshitz
  point,'' {\em Phys. Rev. D}, vol.~79, p.~084008, Apr 2009.

\bibitem{sotiriou2009}
T.~P. Sotiriou, M.~Visser, and S.~Weinfurtner, ``Phenomenologically viable
  lorentz-violating quantum gravity,'' {\em Phys. Rev. Lett.}, vol.~102,
  p.~251601, Jun 2009.

\bibitem{noncomfield}
S.~M. Carroll, J.~A. Harvey, V.~A. Kosteleck\'y, C.~D. Lane, and T.~Okamoto,
  ``Noncommutative field theory and lorentz violation,'' {\em Phys. Rev.
  Lett.}, vol.~87, p.~141601, Sep 2001.

\bibitem{kost1995}
V.~A. Kosteleck\'y and R.~Potting, ``\textit{CPT} , strings, and meson
  factories,'' {\em Phys. Rev. D}, vol.~51, pp.~3923--3935, Apr 1995.

\bibitem{datatables}
V.~A. Kosteleck\'y and N.~Russell, ``Data tables for lorentz and $cpt$
  violation,'' {\em Rev. Mod. Phys.}, vol.~83, pp.~11--31, Mar 2011.

\bibitem{essen1955}
L.~Essen, ``A new aether-drift experiment,'' {\em Nature}, vol.~17,
  pp.~793--794, May 1955.

\bibitem{brillet1979}
A.~Brillet and J.~L. Hall, ``Improved laser test of the isotropy of space,''
  {\em Phys. Rev. Lett.}, vol.~42, pp.~549--552, Feb 1979.

\bibitem{micro1}
J.~A. Lipa, J.~A. Nissen, S.~Wang, D.~A. Stricker, and D.~Avaloff, ``New limit
  on signals of lorentz violation in electrodynamics,'' {\em Phys. Rev. Lett.},
  vol.~90, p.~060403, Feb 2003.

\bibitem{micro4}
P.~L. Stanwix, M.~E. Tobar, P.~Wolf, M.~Susli, C.~R. Locke, E.~N. Ivanov,
  J.~Winterflood, and F.~van Kann, ``Test of lorentz invariance in
  electrodynamics using rotating cryogenic sapphire microwave oscillators,''
  {\em Phys. Rev. Lett.}, vol.~95, p.~040404, Jul 2005.

\bibitem{Schiller2009}
C.~Eisele, A.~Y. Nevsky, and S.~Schiller, ``Laboratory test of the isotropy of
  light propagation at the 10$^{-17}$ level,'' {\em Phys. Rev. Lett.},
  vol.~103, p.~090401, Aug 2009.

\bibitem{optical6}
S.~Herrmann, A.~Senger, K.~M\"ohle, M.~Nagel, E.~V. Kovalchuk, and A.~Peters,
  ``Rotating optical cavity experiment testing lorentz invariance at the
  10$^{-17}$ level,'' {\em Phys. Rev. D}, vol.~80, p.~105011, Nov 2009.

\bibitem{ck2}
D.~Colladay and V.~A. Kosteleck\'y, ``Lorentz-violating extension of the
  standard model,'' {\em Phys. Rev. D}, vol.~58, p.~116002, Oct 1998.

\bibitem{Locke:2008}
C.~R. Locke, E.~N. Ivanov, J.~G. Hartnett, P.~L. Stanwix, and M.~E. Tobar,
  ``Invited article: Design techniques and noise properties of ultrastable
  cryogenically cooled sapphire-dielectric resonator oscillators,'' {\em Rev.
  Sci. Instrum.}, vol.~79, no.~5, p.~051301, 2008.

\bibitem{Kostelecky:2002}
V.~A. Kosteleck\'y and M.~Mewes, ``Signals for lorentz violation in
  electrodynamics,'' {\em Phys. Rev. D}, vol.~66, p.~056005, Sep 2002.

\bibitem{cav1}
P.~L. Stanwix, M.~E. Tobar, P.~Wolf, C.~R. Locke, and E.~N. Ivanov, ``Improved
  test of lorentz invariance in electrodynamics using rotating cryogenic
  sapphire oscillators,'' {\em Phys. Rev. D}, vol.~74, p.~081101, Oct 2006.

\bibitem{cav2}
M.~A. Hohensee, P.~L. Stanwix, M.~E. Tobar, S.~R. Parker, D.~F. Phillips, and
  R.~L. Walsworth, ``Improved constraints on isotropic shift and anisotropies
  of the speed of light using rotating cryogenic sapphire oscillators,'' {\em
  Phys. Rev. D}, vol.~82, p.~076001, Oct 2010.

\bibitem{spinres}
K.~Benmessai, W.~G. Farr, D.~L. Creedon, Y.~Reshitnyk, J.-M. Le~Floch, T.~Duty,
  and M.~E. Tobar, ``Hybrid electron spin resonance and whispering gallery mode
  resonance spectroscopy of fe${}^{3+}$ in sapphire,'' {\em Phys. Rev. B},
  vol.~87, p.~094412, Mar 2013.

\bibitem{bire3}
V.~A. Kosteleck\'y and M.~Mewes, ``Sensitive polarimetric search for relativity
  violations in gamma-ray bursts,'' {\em Phys. Rev. Lett.}, vol.~97, p.~140401,
  Oct 2006.

\bibitem{electrons2003}
H.~M\"uller, S.~Herrmann, A.~Saenz, A.~Peters, and C.~L\"ammerzahl, ``Optical
  cavity tests of lorentz invariance for the electron,'' {\em Phys. Rev. D},
  vol.~68, p.~116006, Dec 2003.

\bibitem{electrons2003-2}
H.~M\"uller, C.~Braxmaier, S.~Herrmann, A.~Peters, and C.~L\"ammerzahl,
  ``Electromagnetic cavities and lorentz invariance violation,'' {\em Phys.
  Rev. D}, vol.~67, p.~056006, Mar 2003.

\bibitem{electrons2005}
H.~M\"uller, ``Testing lorentz invariance by the use of vacuum and matter
  filled cavity resonators,'' {\em Phys. Rev. D}, vol.~71, p.~045004, Feb 2005.

\bibitem{prutt15}
T.~Pruttivarasin, M.~Ramm, S.~G. Porsev, I.~I. Tupitsyn, M.~S. Safronova, M.~A.
  Hohensee, and H.~Haffner, ``Michelson-{Morley} analogue for electrons using
  trapped ions to test {Lorentz} symmetry,'' {\em Nature}, vol.~517,
  pp.~592--595, Jan. 2015.

\bibitem{Mewes2012}
M.~Mewes, ``Optical-cavity tests of higher-order lorentz violation,'' {\em
  Phys. Rev. D}, vol.~85, p.~116012, Jun 2012.

\bibitem{nand2013}
N.~R. Nand, S.~R. Parker, E.~N. Ivanov, J.-M. le~Floch, J.~G. Hartnett, and
  M.~E. Tobar, ``Resonator power to frequency conversion in a cryogenic
  sapphire oscillator,'' {\em Applied Physics Letters}, vol.~103, no.~4,
  pp.~--, 2013.

\bibitem{parkerhigherorder}
S.~R. Parker, M.~Mewes, P.~L. Stanwix, and M.~E. Tobar, ``Cavity bounds on
  higher-order lorentz-violating coefficients,'' {\em Phys. Rev. Lett.},
  vol.~106, p.~180401, May 2011.

\bibitem{cav7}
H.~M\"uller, P.~L. Stanwix, M.~E. Tobar, E.~Ivanov, P.~Wolf, S.~Herrmann,
  A.~Senger, E.~Kovalchuk, and A.~Peters, ``Tests of relativity by
  complementary rotating michelson-morley experiments,'' {\em Phys. Rev.
  Lett.}, vol.~99, p.~050401, Jul 2007.
\end{thebibliography}

\clearpage

\begin{thebibliography}{10}

\bibitem{1881}
Michelson,~A.~A. On the relative motion of the earth and the luminiferous
  ether,'' {\em Am. J. Sci.}~\textbf{22,}~120--129 (1881).

\bibitem{1887}
Michelson,~A.~A. $\&$ Morley,~E.~W. On the relative motion of the earth and the
  luminiferous ether, {\em Am. J. Sci.}~\textbf{34,}~333--345 (1887).

\bibitem{1904}
Morley,~E.~W. $\&$ Miller,~D.~C. Report of an experiment to detect the
  fitzgerald-lorentz effect, {\em Philosophical Magazine}~\textbf{9,}~680--685 (1905).

\bibitem{1924}
Tomaschek,~R. Über das verhalten des lichtes außerirdischer lichtquellen,
  {\em Annalen der Physik}~\textbf{378,}~105--126 (1924).

\bibitem{1926a}
Miller,~D.~C. Significance of the ether-drift experiments of 1925 at mount
  wilson, {\em Science}~\textbf{63,}~433--443 (1926).

\bibitem{1926b}
Kennedy,~R.~J. A refinement of the michelson-morley experiment, {\em Proc.
  Natl. Acad. Sci. USA.}~\textbf{12,}~621--629 (1926).

\bibitem{1927}
Illingworth,~K.~K. A repetition of the michelson-morley experiment using
  kennedy's refinement, {\em Phys. Rev.}~\textbf{30,}~692--696 (1927).

\bibitem{1930}
Joos,~G. Die jenaer wiederholung des michelsonversuchs, {\em Annalen der
  Physik}~\textbf{399,}~385--407 (1930).

\bibitem{1955}
Essen,~L. A new aether-drift experiment,'' {\em Nature}~\textbf{17,}~793--794 (1955).

\bibitem{1964}
Jaseja,~T.~S., Javan,~A., Murray,~J. $\&$ Townes,~C.~H. Test of special
  relativity or of the isotropy of space by use of infrared masers,'' {\em
  Phys. Rev.}~\textbf{133,}~1221--1225 (1964).

\bibitem{1969}
Shamir,~J. $\&$ Fox,~R. A new experimental test of special relativity,'' {\em
  Il Nuovo Cimento B Series 10}~\textbf{62,}~258--264 (1969).

\bibitem{1979}
Brillet,~A. $\&$ Hall,~J.~L. Improved laser test of the isotropy of space,
  {\em Phys. Rev. Lett.}~\textbf{42,}~549--552 (1979).

\bibitem{2003a}
Lipa,~J.~A., Nissen,~J.~A., Wang,~S., Stricker,~D.~A. $\&$ Avaloff,~D. New limit
  on signals of lorentz violation in electrodynamics, {\em Phys. Rev. Lett.}~\textbf{90,}~060403 (2003).

\bibitem{2003b}
Wolf,~P. et al. Tests of lorentz invariance using a microwave resonator, {\em Phys. Rev.
  Lett.}~\textbf{90,}~060402 (2003).

\bibitem{2003c}
M\"uller,~H., Herrmann,~S., Braxmaier,~C., Schiller,~S. $\&$ Peters,~A. Modern
  michelson-morley experiment using cryogenic optical resonators, {\em Phys.
  Rev. Lett.}~\textbf{91,}~020401 (2003).

\bibitem{2004}
Wolf,~P. et al. Improved test of lorentz invariance in electrodynamics, {\em Phys. Rev.
  D}~\textbf{70,}~051902 (2004).

\bibitem{2005a}
Antonini,~P., Okhapkin,~M., G\"okl\"u,~E. $\&$ Schiller,~S. Test of constancy of
  speed of light with rotating cryogenic optical resonators, {\em Phys. Rev.
  A}~\textbf{71,}~050101 (2005).

\bibitem{2005b}
Herrmann,~S., Senger,~A., Kovalchuk,~E., M\"uller, H. $\&$ A.~Peters Test of the
  isotropy of the speed of light using a continuously rotating optical
  resonator, {\em Phys. Rev. Lett.}~\textbf{95,}~150401 (2005).

\bibitem{2006}
Stanwix,~P.~L., Tobar,~M.~E., Wolf,~P., Locke,~C.~R. $\&$ Ivanov,~E.N. Improved
  test of lorentz invariance in electrodynamics using rotating cryogenic
  sapphire oscillators, {\em Phys. Rev. D}~\textbf{74,}~081101 (2006).

\bibitem{2009a}
Eisele,~C., Nevsky,~A.~Y. $\&$ Schiller,~S. Laboratory test of the isotropy of
  light propagation at the 10$^{-17}$ level, {\em Phys. Rev. Lett.}~\textbf{103,}~090401 (2009).

\bibitem{2009b}
Herrmann, S. et al. Rotating optical cavity experiment testing lorentz invariance at the
  10$^{-17}$ level, {\em Phys. Rev. D}~\textbf{80,}~105011 (2009).

\end{thebibliography}
\end{document}